\documentstyle[twoside,fleqn,espcrc2,epsf]{article}

\title{QCD hadron spectrum with domain wall fermions}
 
\author{
  Lingling Wu\address{Department of Physics, Columbia University,
  New York, NY 10027, USA }\thanks{This work was done in collaboration
  with T. Blum, P. Chen, N. Christ, M. Creutz, C. Dawson, G. Fleming,
  A. Kaehler, T. Klassen, C. Malureanu, R. Mawhinney, S. Ohta,
  S. Sasaki, G. Siegert, A. Soni, C. Sui, P. Vranas, M. Wingate, Y
  Zhestkov(RIKEN-BNL-CU collaboration). 
  This work was supported in part by the Department of Energy and the
  RIKEN Brookhaven Research Center.
  Presented at Lattice'99. }
}
 
\begin{document}

\def\thepage{CU--TP--951}
\thispagestyle{myheadings}

\begin{abstract}

We present the QCD hadron spectrum for the cases of both quenched and
two-flavor dynamical domain wall fermions. We compare the results obtained
using the Wilson gauge action and a renormalization group improved
gauge action. Finite volume effects and the dependence on the finite 
extent of the fifth dimension are discussed. 

\end{abstract}

\maketitle

\section{INTRODUCTION}

The domain wall fermions (DWF)
formalism\cite{kaplan}\cite{furman-shamir} uses an extra space-time
dimension to separate the chiral limit from the continuum limit. In
this paper, we report a study of the hadron spectrum obtained from
both quenched and dynamical DWF. 

Our conventions are: $m_f$ is the 4-d bare quark mass that explicitly
mixes the two chiralities on the domain walls, $L_s$ is the extent of
lattice in the fifth dimension, and $m_0$ is the 5-d bare
quark mass, which is often called domain wall height. The masses in
physical units in this paper are obtained by using the rho mass to set the
scale. 

\section{QUENCHED QCD SPECTRUM}

\begin{figure}[htb]
\epsfxsize=\hsize
\epsfbox{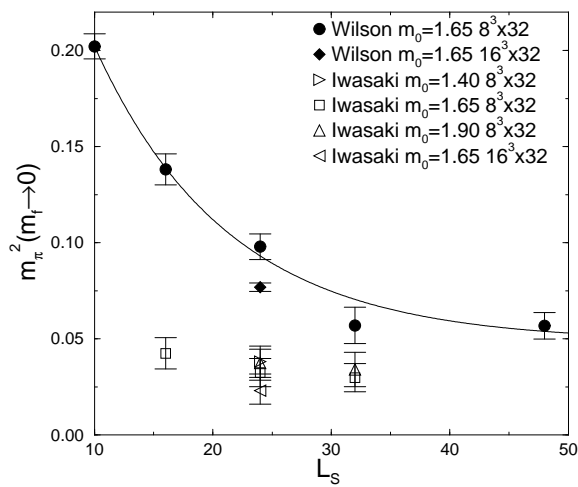}
\vspace{-2pc} 
\caption{
Quenched $m_\pi^2 (m_f \rightarrow 0)$ versus $L_s$ for the Wilson
gauge action ($\beta=5.7$, closed symbols) and the Iwasaki gauge
action ($\beta = 2.2827$, open symbols). The solid line is
$0.048(9) + 0.37(5) \exp ( -0.09(2) L_s )$, with $\chi^2/dof=
2.8/2$. 
}
\label{fig:ls_dep}
\vspace{-1pc}
\end{figure}

Last year, we reported that with domain wall fermions at $\beta = 5.7$
on an $8^3  \times 32$ lattice with $ m_0 = 1.65$ and $ L_s = 48 $, 
$ m_N / m_\rho = 1.42(10) $ as $ m_f \rightarrow 0
$\cite{mawhinney}. Considering the moderate size of the lattice, this
value is favorable compared with Wilson and staggered
results. However, an $L_s$ study for $8^3  \times 32$ lattices at this $\beta$
shows that $m_\pi^2 (m_f \rightarrow 0)$ fits well to the form of $A
\exp ( - \alpha L_s) + B $ with a non-zero value of $B \approx
0.048$, which gives $m_\pi/a (m_f \rightarrow 0) \approx 213 {\rm MeV}$ at
infinite $L_s$\cite{mawhinney}. 
In Figure \ref{fig:ls_dep} we show the pion mass data
on which the conclusion was based.

In order to investigate how much of this $213 {\rm MeV}$ pion 
mass comes from the effects of finite volume, we have
studied a $16^3\times 32$ lattice at $\beta=5.7$, 
$L_s=24$. We find $m_\pi^2(m_f \rightarrow 0)=0.077(2)$, only
$0.021$ smaller than the $0.098(7)$ value
we obtained for $8^3\times 32$ at $L_s=24$. This finite
volume shift is less than half of the $0.048$ infinite 
$L_s$ limit for $8^3\times 32$ suggesting this non-zero
limit of $m_\pi^2$ is not a finite volume effect.  
 
To examine this question more carefully, we assume 
the effects of $L_s$ can be represented by a residual
quark mass, $m_{\rm res}(L_s)$, and express $m_\pi^2$ to 
first order in chiral symmetry breaking as: $m_\pi^2(V,m_f,L_s)=
c_0(V)+c_1(V)*(m_f+m_{\rm res}(L_s))$. We find $c_1(8^3)=4.54(9)$,
$c_1(16^3)=4.75(3)$. If we require that
$c_0$ vanishes as $V\rightarrow \infty$ and assume that
$16^3$ is sufficiently large that $c_0(16^3)\approx0$
then either $m_{\rm res}$ does not vanish as $L_s\rightarrow \infty$ 
or $m_{\rm res}(L_s)$ has a strong, 
unphysical dependence on the spatial volume with 
$m_{\rm res}(24)$ increasing from $0.011(2)$ to $0.0162(5)$  
as the volume is increased from $8^3$ to $16^3$. Thus,
our results appear to require one of three
unexpected possibilities: i) quenched $m_\pi$
does not vanish in the chiral limit; ii) $m_{\rm res}$ does 
not vanish as $L_s \rightarrow \infty$ or iii) $m_{\rm res}$
has a strong volume dependence for $L_s=24$.

\pagenumbering{arabic}
\addtocounter{page}{1}

To support our weak matrix element study\cite{blum}, we have also
measured hadron masses using DWF and the Wilson gauge action at a
weaker coupling, $\beta = 6.0$. On a $16^3 \times 32$ lattice with $m_0 =
1.80$, $L_s = 16$ and $m_f$ ranging from $0.01$ to $0.04$, we
have(Table \ref{val_ex}) $ m_N / m_\rho = 1.37(5) $, $m_\pi^2 =
0.014(2)$ as $ m_f \rightarrow 0 $, which gives 
$m_\pi/a (m_f \rightarrow 0) = 230(15) {\rm MeV}$. 

\begin{table}[htb]
\vspace{0.5pc}
\caption{Valence extrapolations ($a + b m_f^{(val)}$) for $m_\pi^2$,
$m_\rho$ and $m_N$. 
Quenched fits are obtained on a $16^3\times32$ lattice with
$m_0=1.80$. 
Dynamical fits are obtained on an $8^3\times32$ lattice with $m_0=1.90$.}
\label{val_ex}
\begin{tabular}{ccccc}
\hline \hline
mass & $\beta$ & $L_s$ & $a$ & $b$ \\ \hline \hline
\multicolumn{5}{l}{quenched:} \\ \hline 
$m_\pi^2$ & 6.0 & 16 & 0.014(2) & 2.96(7) \\
$m_\rho$  & 6.0 & 16 & 0.400(8) & 2.80(13)  \\
$m_N$     & 6.0 & 16 & 0.55(2)  & 4.7(3)  \\ \hline \hline
\multicolumn{5}{l}{dynamical with $m_f^{(dyn)}=0.02$:} \\ \hline 
$m_\pi^2$ & 5.325 & 24 & 0.325(4) & 5.23(3) \\
$m_\rho$  & 5.325 & 24 & 1.13(2) & 1.80(6) \\
$m_N$     & 5.325 & 24 & 1.62(3) & 3.48(13) \\ \hline
$m_\pi^2$ & 1.90  & 24 & 0.244(3) & 6.07(3) \\ 
$m_\rho$  & 1.90  & 24 & 1.124(11) & 1.90(4) \\
$m_N$     & 1.90  & 24 & 1.66(3) & 3.36(12) \\  \hline
$m_\pi^2$ & 2.0  & 24 & 0.100(6) & 6.31(6) \\
$m_\rho$  & 2.0  & 24 & 0.94(2) & 2.10(6) \\
$m_N$     & 2.0  & 24 & 1.38(2) & 3.65(8) \\ \hline
$m_\pi^2$ & 2.0  & 48 & 0.047(10) & 6.47(11) \\ \hline\hline
\multicolumn{5}{l}{dynamical with $m_f^{(dyn)}=0.06$:} \\ \hline
$m_\pi^2$ & 5.325 & 24 & 0.332(3) & 5.16(2) \\\hline
$m_\pi^2$ & 2.0  & 24 & 0.120(8) & 6.44(8) \\ \hline\hline

\end{tabular}
\vspace{-2pc}
\end{table}  

From the above discussion, we can see that to decrease the pion mass while
using DWF and the Wilson gauge action, large $L_s$ is needed. 
However, the simulation difficulty is proportional to $L_s$.
It might be helpful to use the renormalization group improved gauge
action of Iwasaki which smooths out the gauge fields on the 
lattice\cite{iwasaki}. 

We have simulated using the Iwasaki gauge at $\beta =
2.2827 $ and $c_1 = -0.331$ which is equivalent to quenched $\beta =
5.7 $ with the Wilson action. The resulting pion masses are
promising. Compared with the large $m_\pi^2 (m_f \rightarrow
0)$ values using the Wilson action, much smaller values are obtained
from lattice size $8^3 \times 32$ as shown in Figure \ref{fig:ls_dep}. Studies
with $L_s$ ranging from $16$ to $32$ and $m_0$ ranging from $1.40$ to
$1.90$ show that $m_\pi$ has little dependence on $L_s$ and $m_0$. A
simulation at larger volume ($16^3 \times 32$) also shows that
the finite volume effect is small(Figure \ref{fig:ls_dep}). These
suggest that using the Iwasaki gauge action may enable us to study
quenched DWF at smaller $Ls$, but we may still have a non-zero
pion mass at infinite $L_s$. 

\section{DYNAMICAL QCD SPECTRUM}

To support our thermodynamics studies\cite{vranas99}, we have 
measured the dynamical QCD hadron masses at zero temperature to set
the scale. Unless indicated otherwise, all the simulations discussed
in this section are done with lattice size $8^3 \times 32$, $m_0 =
1.90$, $L_s =24 $, $m_f^{(dyn)} = 0.02$, and valence quark mass ranging
from 0.02 to 0.22 with an increment of 0.04. Valence extrapolations
for some of the simulations discussed below are shown in Table
\ref{val_ex}.  

Using DWF and the Wilson action, at $N_t = 4$ the
transition occurs at about $\beta = 5.325$. At this $\beta$, we find 
$m_\rho = 1.18(3)$, $m_\pi = 0.654(3)$ at $m_f^{(dyn)} = 0.02$,
which gives $T_c = 163(4) {\rm MeV}$ and $m_\pi/a = 427(11) {\rm
MeV}$. Using a larger $16^3 \times 16$ volume only reduces $m_\pi$
to $0.652(3)$, which suggests that the finite volume effect is very small. 
A Ward identity evaluation\cite{gfleming} shows that the residual mass
caused by the mixing between the two walls plays an important role in
this heavy pion mass. However, that study shows these residual mass
effects do vanish in the limit of large $L_s$ for these full QCD simulations.

We have also measured the masses for $m_f^{(dyn)} = 0.06$ at $\beta =
5.325$. Performing real dynamical extrapolations using the two dynamical
points obtained from $m_f=0.02, 0.06$, we get
$m_\pi^2 = 0.320(6) + 5.38(11) m_f^{(dyn)}$.
Compared with valence extrapolation, the dynamical extrapolation slightly
decreases the pion mass as $m_f \rightarrow 0$. 

As in the quenched studies, we have also investigated the
renormalization group improved gauge action. We have simulated at 
$\beta = 1.90$, $c_1 = -0.331$ which is about the transition point for
$N_t = 4$ with DWF and the Iwasaki gauge action\cite{vranas99}. 
We obtain $m_\rho = 1.16(2)$, $m_\pi = 0.604(3)$ at $m_f^{(dyn)} =
0.02$, which gives $T_c = 166(3) {\rm MeV}$ and $m_\pi/a = 400(7) {\rm
MeV}$. Surprisingly this is about the same value as that obtained at
$\beta_{c}$ using the Wilson action. Therefore, although the Iwasaki
action helps to reduce the pion mass in our quenched study, this is
not true for the dynamical case.

We have also measured the hadron masses at $\beta = 2.0$ for both
$m_f^{(dyn)}=0.02, 0.06$. As with the Wilson gauge action, we can draw
the same conclusion that the dynamical extrapolation $m_\pi^2 =
0.088(10) + 6.9(2) m_f^{(dyn)}$ gives a slightly smaller pion mass
as $m_f \rightarrow 0$. 

\begin{figure}[htb]
\vspace{1pc}
\epsfxsize=\hsize
\epsfbox{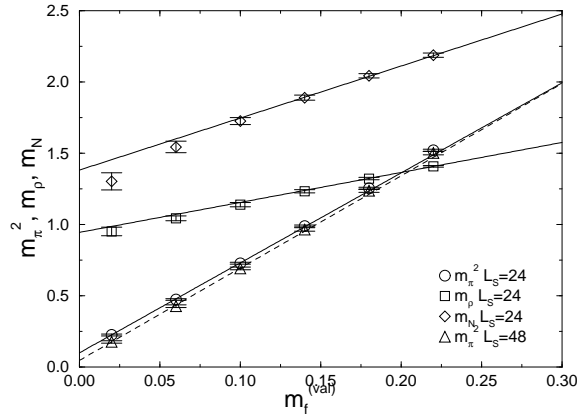}
\caption{
The quantities $m_\pi^2$, $m_\rho$, $m_N$ plotted versus $m_f^{(val)}$ at $\beta = 2.0$, $c_1
= -0.331$, $m_0 = 1.90$, $m_f^{(dyn)} = 0.02$ on $8^3 \times 32$ using
the Iwasaki
action. The solid lines are valence extrapolations for $m_\pi^2$, 
$m_\rho$ and $m_N$ at $L_s =
24$. The dashed line is the valence extrapolation for
$m_\pi^2$ at $L_s =48$. 
}
\label{fig:ls_dyn}
\vspace{-1pc}
\end{figure}

To study the finite $L_s$ effect, we are currently doing a simulation
at $\beta=2.0$, $c_1=-0.331$, and $L_s=48$. Figure \ref{fig:ls_dyn}
shows the valence extrapolations of the pion, rho, nucleon mass values
for $L_s=24$ and the pion mass for $L_s=48$. 
For $L_s =24$, we have $m_\pi (m_f^{(dyn)} = 0.02) = 0.475(7)$.
For $Ls = 48$, we have obtained $m_\pi (m_f^{(dyn)} = 0.02) =
0.420(10)$. This confirms that the mixing between the two walls is at
least a major cause of the heavy pion mass at $\beta_c$, and this
effect can be reduced by increasing $L_s$.

\section{CONCLUSIONS}

Although for the quenched QCD spectrum the Iwasaki action lowers
the pion mass compared with that obtained using the Wilson action, the
non-zero value of $m_\pi (m_f \rightarrow 0)$ is still problematic. 
From our dynamical QCD spectrum study, we have obtained a heavy pion mass at
$\beta_c$ for $N_t = 4$ for both the Wilson and Iwasaki gauge
actions at $L_s = 24$. This can be improved by increasing the size of $L_s$. 
These calculations were performed on the QCDSP machines at Columbia
and RIKEN/BNL.

\end{document}